\documentclass[ aip,jmp, reprint,twocolumn]{revtex4-1}
\usepackage{amssymb}
\usepackage{amsmath}
\usepackage{graphicx}
\usepackage{dcolumn}
\usepackage{bm}
\usepackage{multirow}

\begin{document}

\title{Atom-light hybrid quantum gyroscope}
\author{Yuan Wu}
\affiliation{State Key Laboratory of Precision Spectroscopy, Quantum
Institute of Atom and Light, Department of Physics, East China Normal
University, Shanghai 200062, P. R. China}
\author{Jinxian Guo}
\affiliation{School of Physics and Astronomy, and Tsung-Dao Lee Institute,
Shanghai Jiao Tong University, Shanghai 200240, P. R. China}
\affiliation{Shanghai
Research Center for Quantum Sciences, Shanghai 201315, P. R. China}
\author{Xiaotian Feng}
\affiliation{State Key Laboratory of Precision Spectroscopy, Quantum
Institute of Atom and Light, Department of Physics, East China Normal
University, Shanghai 200062, P. R. China}
\author{L.Q.Chen}
\email{lqchen@phy.ecnu.edu.cn}
\affiliation{State Key Laboratory of Precision Spectroscopy, Quantum
Institute of Atom and Light, Department of Physics, East China Normal
University, Shanghai 200062, P. R. China}
\affiliation{Shanghai Research
Center for Quantum Sciences, Shanghai 201315, P. R. China}
\author{Chun-Hua Yuan}
\email{chyuan@phy.ecnu.edu.cn}
\affiliation{State Key Laboratory of Precision Spectroscopy, Quantum
Institute of Atom and Light, Department of Physics, East China Normal
University, Shanghai 200062, P. R. China}
\author{Weiping Zhang}
\affiliation{School of Physics and Astronomy, and Tsung-Dao Lee Institute,
Shanghai Jiao Tong University, Shanghai 200240, P. R. China}
\affiliation{Shanghai
Research Center for Quantum Sciences, Shanghai 201315, P. R. China}
\affiliation{Collaborative Innovation Center of Extreme Optics, Shanxi
University, Taiyuan, Shanxi 030006, P. R. China}
\date{\today }

\begin{abstract}
A new type of atom-light hybrid quantum gyroscope (ALHQG) is proposed due to
its high rotation sensitivity. It consists of an optical Sagnac loop to
couple rotation rate and an atomic ensemble as quantum beam
splitter/recombiner (QBS/C) based on atomic Raman amplification process to
realize the splitting and recombination of the optical wave and the atomic
spin wave. The rotation sensitivity can be enhanced by the quantum
correlation between Sagnac loop and QBS/C. The optimal working condition
is investigated to achieve the best sensitivity. The numerical results show
that the rotation sensitivity can beat the standard quantum limit (SQL) in
ideal condition. Even in the presence of the attenuation under practical
condition, the best sensitivity of the ALHQG can still beat the SQL and is
better than that of a fiber optic gyroscope (FOG). Such an ALHQG could be
practically applied for modern inertial navigation system.
\end{abstract}

\pacs{Valid PACS appear here}
\maketitle




\section{Introduction}

Highly accurate and precise rotation measuring instruments are fundamental
apparatus in inertial navigation, geophysical studies and tests of general
relativity \cite{1}.\textbf{\ }Rotation sensors based on the Sagnac effect
\cite{2} have been constructed using light-wave and matter-wave (neutrons,
neutral atoms and electrons). The Sagnac phase \cite{3,4} is\ caused by an
interferometer rotating at rate $\Omega $ which is related with the velocity
of the particle $\nu $, the loop area $A$ and the wavelength $\lambda $.
Regarding the matter-wave gyroscope, such as the atomic gyroscope \cite{5},
it has large rotation sensitivity per unit area \cite{6,7} and realizes high
rotation sensitivity \cite{8}. However, it possess a small bandwidth and
suffers from low repetition rate and dead times during which no inertial
measurement can be made \cite{9}. The light-wave gyroscope, such as FOG \cite%
{10}, has large loop area, simple system and also realizes high sensitivity,
but its rotation sensitivity is limited by the SQL \cite{11}. The
limitations of the matter-wave and light-wave gyroscopes affect their
practical application and further performance improvement.

To improve the performance of the matter-wave and light-wave gyroscopes,
some hybrid strategies have been reported. One strategy is based on the
combination of the mechanical sensor and the atomic sensor \cite{12,13} to
overcome the limitations of low bandwidth and the dead time issue in atomic
sensor. Other strategy, such as the electromagnetically induced transparency
(EIT) in cold atomic system, has been proposed to realize the associated
momentum transfer from light to atom to enhance the sensitivity of
light-wave sensor. However, the sensitivity of above hybrid strategies were
limited by the SQL \cite{13}. Because there are no quantum correlation in
Sagnac loop or hybrid sensors.

Recently, some nonlinear effects have been proposed that can break through
SQL to enhance the sensitivity. In 2017, a new nonlinear Sagnac rotation
sensor based on four-wave mixing (FWM) was proposed \cite{16}. Such a sensor
can beat the SQL in an ideal case due to quantum correlation between two
Sagnac beams, while\ its practical situation has been poorly discussed.
Furthermore, it is difficult to realize the Sagnac loop and phase
stabilization for four beams \cite{17}. Currently, a new type of hybrid
atom-light interferometer \cite{18} has been demonstrated where Raman
amplification processes in atomic ensemble act as QBS/C of optical wave and
atomic system. The quantum correlation between optical wave and atomic
ensemble leads to a high-contrast interference fringe. The phase sensitivity
of the interferometer can beat the SQL by the factor of the amplification
gain of the QBS/C in principle \cite{19}.

In this work, an atom-light hybrid quantum gyroscope (ALHQG) is proposed. It
is an optical Sagnac interferometer with the beam splitter/recombiner
replaced by QBS/C to realize the quantum correlation between the optical
wave and the atomic spin wave. The rotation sensitivity is analyzed with
practical parameters in real experiment, including the particle number of
the input field, the gain of the Raman-amplification process, the Sagnac
fiber loop length, the\ attenuation coefficient of photon and atom, etc. It
is found that, due to quantum correlation, the rotation sensitivity of the
proposed gyroscope can beat the SQL in an ideal case. Even if the optical
loss and atomic decoherence are considered in the ALHQG, the sensitivity can
still beat the SQL and is better than the FOG\textbf{\ }with the same
rotation-sensitive particle number. Such an ALHQG has significantly
practical value in quantum metrology.

The paper is organized as follow: in Sec. \ref{sec2}, the working principle
of the ALHQG in practice is described. In Sec. \ref{sec3}, the sensitivity
of the ALHQG is analysed to have optimal working condition. In Sec. \ref%
{sec4}, the intensity and the frequency fluctuation of laser are analyzed
under optimal working condition. In Sec. \ref{sec5}, a summary of our
results is concluded.

\section{Theory and principles of ALHQG \label{sec2}}

\begin{figure}[tbp]
\centering
\includegraphics[scale=0.38]{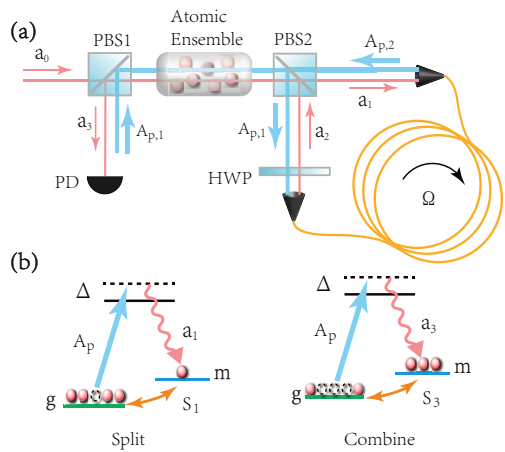}
\caption{(a) Schematic of the ALHQG. Red: Stokes field; blue: Raman write
field; PBS: polarization beam splitter; HWP: half wave plate to rotate the
polarization angles by 90 degrees; FC: fiber coupler; SMF: single-mode
fiber; PD: photo-detectors. (b) The energy levels of atom. The lower two
energy states $\left\vert g\right\rangle $ and $\left\vert m\right\rangle $
are the hyperfine split ground states. The higher energy states $\left\vert
e\right\rangle $ are the excited states. $\Delta $ is the single photon
detuning. A strong Raman write beam $A_{p,1}$ ($A_{p,2}$ ) couples the state
$\left\vert e\right\rangle $ with $\left\vert g\right\rangle $ and generates
a Stokes field $\hat{a}_{1}$ ($\hat{a}_{3}$) and the corresponding atomic
spin wave $\hat{S}_{1}$ ($\hat{S}_{3}$). The atomic spin wave stays in the
cell, and the Stokes field travels out together with the pump field.}
\label{fig:Fig1}
\end{figure}

The scheme of the ALHQG is shown in Fig.~\ref{fig:Fig1} (a). It consists of
an optical Sagnac loop to couple rotation rate $\Omega $ and an atomic
ensemble as QBS/C to generate the quantum correlated optical and atomic
waves and then recombine the waves for interference. The energy levels of
atom are given in Fig. \ref{fig:Fig1}(b). A strong Raman write beam $A_{p,1}$
and a weak Stokes input field $\hat{a}_{0}$ with orthogonal polarizations
interact with a $\Lambda $-shaped atomic ensemble to generate an amplified
Stokes field $\hat{a}_{1}$ and a correlated atomic spin wave $\hat{S}_{1}$
via the first Raman amplification.\textbf{\ }The optical field $\hat{a}_{1}$
and atomic spin wave $\hat{S}_{1}$\textbf{\ }have quantum correlation\ and
the relative intensity fluctuations are squeezed \cite{19}. After
interaction, the atomic spin wave $\hat{S}_{1}$ stays in the atomic ensemble
while the Stokes field $\hat{a}_{1}$ and the strong Raman write beam $%
A_{p,1} $ travel together out of the atomic ensemble and propagate in the
opposite directions inside a fiber Sagnac loop. As a result, the lights in
clockwise (CW) and counter-clockwise (CCW) experience a Sagnac phase induced
by the rotation rate $\Omega $. Then the Stokes field $\hat{a}_{2}$ and the
strong Raman write beam $A_{p,2}$ recombine with the waiting atomic spin
wave $\hat{S}_{2} $, evolved from $\hat{S}_{1}$, in the atomic ensemble%
\textbf{\ }to realize the interference between optical wave $\hat{a}_{2}$\
and atomic wave $\hat{S}_{2}$\ via the second Raman amplification. Finally,
the Stokes field $\hat{a}_{3}$ and correlated atomic spin wave $\hat{S}_{3}$
are generated. Therefore, the realization of ALHQG needs three steps, which
are atom-light beam splitting via the first Raman process, Sagnac effect and
atom-light beam combination via the second Raman process to achieve the
rotation rate $\Omega $.

In general, the input-output relation for the atom-light beam splitting via
Raman process is written as \cite{17}:
\begin{eqnarray}
\widehat{a}_{1} &=&G_{1}\widehat{a}_{0}+e^{i\theta _{1}}g_{1}\hat{S}%
_{0}^{\dag }, \\
\widehat{S}_{1} &=&e^{i\theta _{1}}g_{1}\hat{a}_{0}^{\dag }+G_{1}\widehat{S}%
_{0},
\end{eqnarray}%
Here $e^{i\theta _{1}}=\eta A_{p,1}/\left\vert \eta A_{p,1}\right\vert $,
where $\eta $\ is the coupling constant. $G_{1}=\cosh (\left\vert \eta
A_{p,1}\right\vert t)$ and $g_{1}=\sinh (\left\vert \eta A_{p,1}\right\vert
t)$ are the amplitude gains of the first Raman-amplification process,
satisfying $|G_{1}|^{2}-|g_{1}|^{2}=1$. Here $t$ is the pulse duration of
pump field. $\widehat{a}_{0}$\ and $A_{p,1}$\ are initial input Stokes beam
and the Raman write beam. When a\ coherent state $\hat{a}_{0}$\ enters into
the gyroscope and the spin wave $\widehat{S}_{0}$ is initially in a vacuum
state, the particle number of the input Stokes field in a single shot is $%
\langle \hat{a}_{0}^{\dag }\hat{a}_{0}\rangle \equiv N_{in}$. And thus, the
total particle number $N_{Tot}$\ inside the ALHQG include not only the
photon number but also the atomic collective excitation number, which is:
\begin{eqnarray}
N_{Tot} &=&\langle \hat{a}_{1}^{\dag }\hat{a}_{1}\rangle +\langle \hat{S}%
_{1}^{\dag }\hat{S}_{1}\rangle \\
&=&g_{1}^{2}(1+N_{in})+g_{1}^{2}+G_{1}^{2}N_{in},  \notag
\end{eqnarray}%
Here $\left\langle {}\right\rangle $ is a quantum expectation value. When $%
N_{in}\gg 1$, $N_{Tot}\approx (g_{1}^{2}+G_{1}^{2})N_{in}$.

After the beam splitting process,\ the optical wave $\widehat{a}_{1}$\ and
the Raman write beam $A_{p,1}$\ transfer out of the atomic ensemble and
enter the Sagnac loop to be subject to the phase ($\varphi _{cw}$\ and $%
\varphi _{ccw}$) induced by the rotation $\Omega $. Here $\varphi _{cw}$ and
$\varphi _{ccw}$ are the phases of the CW and CCW induced by rotation rate,
respectively.\ Under the influence of the optical fiber loss and the\ atomic
decoherence,\ the input-output relation in the Sagnac loop is:
\begin{eqnarray}
\widehat{a}_{2} &=&\sqrt{T}\widehat{a}_{1}e^{i\varphi _{cw}}+\sqrt{R}%
\widehat{V}_{cw},  \label{A} \\
A_{p,2} &=&\sqrt{T}e^{i\varphi _{ccw}}A_{p,1}, \\
\widehat{S}_{2} &=&\widehat{S}_{1}e^{-\Gamma \tau }+\widehat{F},  \label{B}
\end{eqnarray}%
where $T=\exp (-\alpha _{T}L)$ and $R=1-T$ are the transmission and
reflectance coefficients of the photons, respectively. Here $\alpha _{T}$ is
the fiber attenuation coefficient and $L$ is the length of fiber loop. $%
\widehat{V}_{cw}$ is the operator of the vacuum. $e^{-\Gamma \tau }$ is the
collisional dephasing of atomic excitation and $\Gamma $ is the
corresponding decay rate. $\widehat{F}$ is the Langevin operator and
satisfies $\langle \widehat{F}\widehat{F}^{\dagger }\rangle =1-e^{-2\Gamma
\tau }$.

Then, $\widehat{a}_{2}$ and $A_{p,2}$\ recombine with the correlated $%
\widehat{S}_{2}$\ via the second Raman amplification to obtain the final
outputs $\widehat{a}_{3}$\ and $\widehat{S}_{3}$, which are:%
\begin{eqnarray}
\widehat{a}_{3} &=&G_{2}\widehat{a}_{2}+e^{i(\theta _{2}+\varphi
_{ccw})}g_{2}\hat{S}_{2}^{\dag }, \\
\widehat{S}_{3} &=&e^{i(\theta _{2}+\varphi _{ccw})}g_{2}\hat{a}_{2}^{\dag
}+G_{2}\widehat{S}_{2},
\end{eqnarray}%
where $e^{i(\theta _{2}+\varphi _{ccw})}=\eta A_{p,2}/\left\vert \eta
A_{p,2}\right\vert $. $G_{2}=\cosh (\left\vert \eta A_{p,2}\right\vert t)$
and $g_{2}=\sinh (\left\vert \eta A_{p,2}\right\vert t)$ are the amplitude
gains of the second Raman amplification process, which also satisfy $%
|G_{2}|^{2}-|g_{2}|^{2}=1$. And thus, the final outputs are:%
\begin{eqnarray}
\widehat{a}_{3} &=&A_{1}\hat{a}_{0}+B_{1}\hat{S}_{0}^{\dag }+C_{1}\hat{V}%
_{cw}+D_{1}\hat{F}^{\dagger }, \\
\widehat{S}_{3} &=&A_{2}\hat{a}_{0}^{\dag }+B_{2}\hat{S}_{0}+C_{2}\hat{V}%
_{cw}^{\dag }+D_{2}\hat{F},
\end{eqnarray}%
where
\begin{eqnarray*}
A_{1} &=&\sqrt{T}G_{1}G_{2}e^{i\varphi _{cw}}+g_{1}^{\ast }g_{2}e^{-\Gamma
\tau }e^{i(\varphi _{ccw}+\theta _{2}-\theta _{1})}, \\
B_{1} &=&\sqrt{T}g_{1}G_{2}e^{i(\varphi _{cw}+\theta _{1})}+G_{1}^{\ast
}g_{2}e^{-\Gamma \tau }e^{i(\varphi _{ccw}+\theta _{2})}, \\
C_{1} &=&\sqrt{R}G_{2},\text{ }D_{1}=g_{2}e^{i(\varphi _{ccw}+\theta _{2})},
\\
A_{2} &=&\sqrt{T}G_{1}^{\ast }g_{2}e^{i(\theta _{2}+\varphi _{ccw}-\varphi
_{cw})}+g_{1}G_{2}e^{-\Gamma \tau }e^{i\theta _{1}}, \\
B_{2} &=&\sqrt{T}g_{1}^{\ast }g_{2}e^{i(\theta _{2}-\theta _{1}+\varphi
_{ccw}-\varphi _{cw})}+G_{1}G_{2}e^{-\Gamma \tau }, \\
C_{2} &=&\sqrt{R}g_{2}e^{i(\theta _{2}+\varphi _{ccw})},\text{ }D_{2}=G_{2}.
\end{eqnarray*}

Normally, there are two detection methods, homodyne detection and intensity
detection, to detect the output and obtain the rotation rate $\Omega $. In
experiment, the operation of intensity detection is simpler. And thus,%
\textbf{\ }the photon number operator $\left\langle \widehat{n}\right\rangle
=\left\langle \widehat{a}_{3}^{\dag }\widehat{a}_{3}\right\rangle $ is
employed as the measurable operator in intensity detection, which is:
\begin{equation}
\left\langle \widehat{n}\right\rangle =\left\vert A_{1}\right\vert
^{2}N_{in}+\left\vert B_{1}\right\vert ^{2},
\end{equation}%
where $\left\vert A_{1}\right\vert ^{2}=T\left\vert G_{1}\right\vert
^{2}\left\vert G_{2}\right\vert ^{2}+\left\vert g_{1}\right\vert
^{2}\left\vert g_{2}\right\vert ^{2}e^{-2\Gamma \tau }+2\sqrt{T}\left\vert
G_{1}\right\vert \left\vert G_{2}\right\vert \left\vert g_{1}\right\vert
\left\vert g_{2}\right\vert e^{-\Gamma \tau }\cos \left( \beta \Omega
+\theta _{1}-\theta _{2}\right) $, $\left\vert B_{1}\right\vert
^{2}=T\left\vert g_{1}\right\vert ^{2}\left\vert G_{2}\right\vert
^{2}+\left\vert G_{1}\right\vert ^{2}\left\vert g_{2}\right\vert
^{2}e^{-2\Gamma \tau }+2\sqrt{T}\left\vert G_{1}\right\vert \left\vert
G_{2}\right\vert \left\vert g_{1}\right\vert \left\vert g_{2}\right\vert
e^{-\Gamma \tau }\cos \left( \beta \Omega +\theta _{1}-\theta _{2}\right) $.
Here $\beta \Omega =\varphi _{cw}-\varphi _{ccw}$ is the Sagnac phase, where
$\beta =2\pi DL/(\lambda c)$, $D$\ is the diameter of the Sagnac loop, $L$
is the length of the Sagnac loop and $c$\ is the speed of light in vacuum.

Based on the error-propagation analysis \cite{20}, the rotation sensitivity $%
\Delta \Omega $ in a single shot is defined as:
\begin{equation}
\Delta \Omega =\frac{\langle (\Delta \widehat{n})^{2}\rangle ^{1/2}}{%
\left\vert \partial \langle \widehat{n}\rangle /\partial \Omega \right\vert }%
,  \label{RSN}
\end{equation}%
here $\langle (\Delta \widehat{n})^{2}\rangle =\langle \widehat{n}%
^{2}\rangle -\langle \widehat{n}\rangle ^{2}$. The uncertainty $\langle
(\Delta \widehat{n})^{2}\rangle $ and the slope $\left\vert \partial \langle
\widehat{n}\rangle /\partial \Omega \right\vert $ are respectively given by:
\begin{eqnarray}
\langle (\Delta \widehat{n})^{2}\rangle &=&\left\vert A_{1}\right\vert
^{4}N_{in}+\left( \left\vert A_{1}\right\vert ^{2}N_{in}+\left\vert
B_{1}\right\vert ^{2}\right) \left\vert C_{1}\right\vert ^{2}  \notag \\
&&+[\left\vert A_{1}\right\vert ^{2}(1+N_{in})+\left\vert C_{1}\right\vert
^{2}]\left\vert D_{1}\right\vert ^{2}(1-e^{-2\Gamma \tau })  \notag \\
&&+\left\vert A_{1}\right\vert ^{2}\left\vert B_{1}\right\vert ^{2}\left(
1+N_{in}\right) ,
\end{eqnarray}%
\begin{equation}
\left\vert \frac{\partial \langle \widehat{n}\rangle }{\partial \Omega }%
\right\vert =2\sqrt{T}\beta \left\vert G_{1}\right\vert \left\vert
G_{2}\right\vert \left\vert g_{1}\right\vert \left\vert g_{2}\right\vert
e^{-\Gamma \tau }\left\vert \sin (\beta \Omega )\right\vert (N_{in}+1).
\end{equation}

When $\theta _{1}-\theta _{2}=\pi $, we have $\left\vert A_{1}\right\vert
^{2}=T\left\vert G_{1}\right\vert ^{2}\left\vert G_{2}\right\vert
^{2}+\left\vert g_{1}\right\vert ^{2}\left\vert g_{2}\right\vert
^{2}e^{-2\Gamma \tau }-2\sqrt{T}\left\vert G_{1}\right\vert \left\vert
G_{2}\right\vert \left\vert g_{1}\right\vert \left\vert g_{2}\right\vert
e^{-\Gamma \tau }\cos \left( \beta \Omega \right) $, $\left\vert
B_{1}\right\vert ^{2}=T\left\vert g_{1}\right\vert ^{2}\left\vert
G_{2}\right\vert ^{2}+\left\vert G_{1}\right\vert ^{2}\left\vert
g_{2}\right\vert ^{2}e^{-2\Gamma \tau }-2\sqrt{T}\left\vert G_{1}\right\vert
\left\vert G_{2}\right\vert \left\vert g_{1}\right\vert \left\vert
g_{2}\right\vert e^{-\Gamma \tau }\cos \left( \beta \Omega \right) $, $%
\left\vert C_{1}\right\vert ^{2}=RG_{2}^{2}$, and $\left\vert
D_{1}\right\vert ^{2}=g_{2}^{2}$. It can be seen that the sensitivity is
related with input particle number ($N_{in}$), gains ($G_{1}$, $G_{2}$, $%
g_{1}$ and $g_{2}$), loop length ($L$), rotation rate ($\Omega $), photon
loss coefficient ($\alpha _{T}$) and atomic decoherence rate ($\Gamma $). To
obtain the best sensitivity, the discussion in the next section is focused
on the optimal working condition of ALHQG.

\section{Sensitivity analysis \label{sec3}}

Based on above analysis, the sensitivity is related with several parameters.
To simplify the analysis, we start by the sensitivity under the ideal
condition, $G_{2}=G_{1}=G$, $g_{2}=g_{1}=g$, $\theta _{1}-\theta _{2}=\pi $,
$T=1$, $R=0$ and\ $e^{-\Gamma \tau }=1$. With $N_{in}\gg 1$, based on Eq. (%
\ref{RSN}), the rotation sensitivity $\Delta \Omega $ in a single shot is
given by
\begin{equation}
\Delta \Omega \approx \frac{1}{M}\frac{1}{\beta \sqrt{N_{in}}},  \label{RS}
\end{equation}%
with
\begin{equation*}
M=\frac{2\left\vert Gg\right\vert ^{2}\left\vert \sin \left( \beta \Omega
\right) \right\vert }{\sqrt{\{1+2\left\vert Gg\right\vert ^{2}[1-\cos (\beta
\Omega )]\}\{1+4\left\vert Gg\right\vert ^{2}[1-\cos (\beta \Omega )]\}}}.
\end{equation*}%
It can be seen that the rotation sensitivity $\Delta \Omega $\ of ALHQG is
inversely proportional to $M$ and $\sqrt{N_{in}}$. $1/(\beta \sqrt{N_{in}})$ is the SQL of the
traditional gyroscope when the input
particle number is $N_{in}$. Hence, based on
Eq. (\ref{RS}), we can see that, with the same input particle number $N_{in}$%
, the ALHQG is\ enhanced by $1/M$ ($M>1$ when $G>1$ and $%
\cos \left( \beta \Omega \right) $\textbf{\ }$\rightarrow 1$) compared with
the traditional gyroscope, such as FOG or ring-laser gyroscope. This is due to the quantum correlation
between $\widehat{a}_{1}$\ and $\widehat{S}_{1}$\ in first Raman process, so
that the signal to noise ratio is enhanced in the second Raman amplification
\cite{19}.

Normally, to ensure a fair comparison, the particle number in SQL should be the rotation-sensitive particle number, which is $N_{Tot}$ in the ALHQG. Then the corresponding SQL is $\Delta \Omega
_{SQL}=$ $1/(\beta \sqrt{N_{Tot}})$ and the sensitivity enhancement factor $K
$\ is:
\begin{equation}
K=\frac{\Delta \Omega }{\Delta \Omega _{SQL}}\simeq \frac{\sqrt{g^{2}+G^{2}}%
}{M},  \label{K1}
\end{equation}%
It can be seen that $K$\ depends only on $G$ and $\beta \Omega $ but not $%
N_{in}$. When $K<1$, the sensitivity of ALHQG can beat the SQL. Fig. 2 shows
$K$\ as a function of $\beta \Omega $\ at different gains $G$. In general,
the enhancement factor $K$\ firstly\ decreases to a minimum value and then
gradually increase with $\beta \Omega $.\textbf{\ }The sensitivity of ALHQG
is always below the SQL at the optimal Sagnac phase marked by pink rhombus
in each curve and denoted by $\Lambda (N_{in},G)$. Meanwhile, the larger $G$
is, the smaller $K$ or the better rotation sensitivity $\Delta \Omega $ can
be obtained. Furthermore, the optimal $\beta \Omega $\ is closer to zero,
which means that the rotation measurement is very sensitive to the change of
the rotation rate. The enhancement of the sensitivity is accompanied with
the decrease of the dynamic range of rotation measurement. In the future
practical application, a balance between sensitivity and dynamic range
should be considered.
\begin{figure}[tbp]
\centering
\includegraphics[scale=0.32]{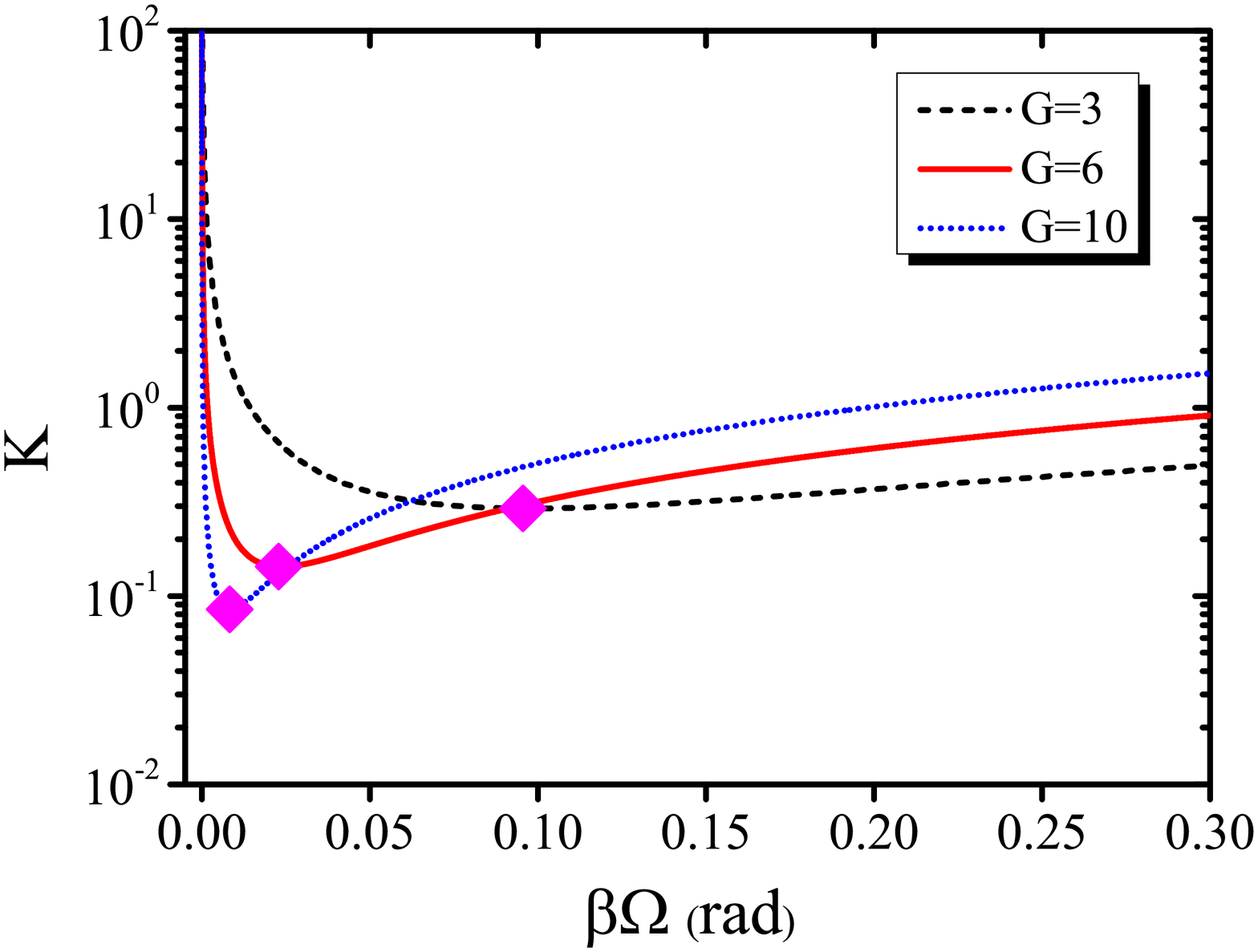}
\caption{The sensitivity enhancement factor $K$\ versus Sagnac phase $%
\protect\beta \Omega $ with different gains $G$ when $N_{in}=10^{8}$ in a
single shot.}
\label{fig:Fig2}
\end{figure}

\begin{figure}[tbp]
\centering
\includegraphics[scale=0.32]{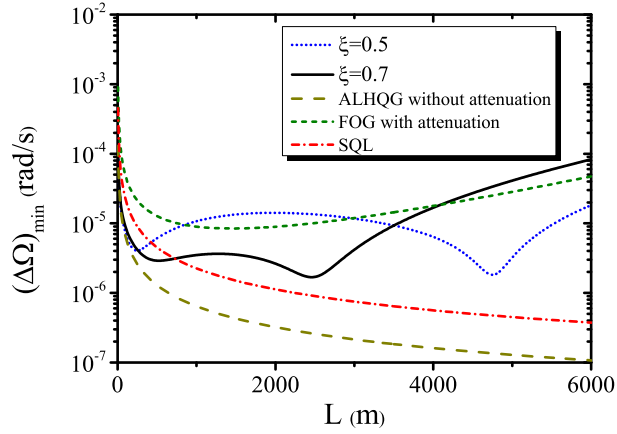}
\caption{The minimum $\Delta \Omega $ versus loop length $L$ with different
attenuation coefficient ratios $\protect\xi $ when $G_{1}=6$, $N_{in}=10^{8}$
in a single shot and $\protect\alpha _{T}=3$ dB/km.}
\label{fig:Fig3}
\end{figure}

Based on above analysis, when $\beta \Omega $ is at the optimal point $%
\Lambda (N_{in},G)$, we have $1-\cos \left( \beta \Omega \right) \mathbf{\ }%
\approx 0$ and $\sin \left( \beta \Omega \right) \approx \Lambda (N_{in},G)$%
. Due to $\beta =2\pi DL/(\lambda c)$, the minimum rotation sensitivity $%
(\Delta \Omega )_{\min }$ is:
\begin{equation}
(\Delta \Omega )_{\min }\approx \frac{\lambda c}{4\pi DL\sqrt{N_{in}}%
\left\vert Gg\right\vert ^{2}\Lambda (N_{in},G)}.  \label{eq17}
\end{equation}%
where $L$ is Sagnac loop length and $\lambda $ is the wavelength. It can be
seen that the larger $N_{in}$, $G$ and $L$ are, the better $(\Delta \Omega
)_{\min }$. As shown in Fig. \ref{fig:Fig3}, $(\Delta \Omega )_{\min }$ in
dark-yellow dash line,\ can always beat the SQL illustrated by the red
dash-dot line.\

However, in practice, the photon loss and atomic decoherence in the ALHQG
can not be ignored. Based on Eq. (\ref{A}), the photon number of the Stokes
light is $\langle \hat{a}_{2}^{\dag }\hat{a}_{2}\rangle =e^{-\alpha
_{T}L}\langle \hat{a}_{1}^{\dag }\hat{a}_{1}\rangle $. The photon number
decreases with the loop length $L$ at the rate of $\alpha _{T}$. At the same
time, based on Eq. (\ref{B}), the atomic number of the spin wave is $\langle
\widehat{S}_{2}^{\dag }\widehat{S}_{2}\rangle \simeq e^{-2\Gamma \tau
}\langle \widehat{S}_{1}^{\dag }\widehat{S}_{1}\rangle $. The atomic number
also decreases with $L$ at the rate of $2n\Gamma /c$ since $\tau =nL/c $
where $n$ is the refractive index of the optical fiber. It is known that the
attenuation leads to poor sensitivity due to the weaker quantum correlation
\cite{19}. Furthermore, due to the different attenuation rates of the light
field ($\alpha _{T}$) and atomic spin wave ($2n\Gamma /c$), the particle
number of the interference arms in ALHQG are unequal. And thus, it is
complex to analyze the dependence of the minimum rotation sensitivity $%
(\Delta \Omega )_{\min }$ on the loop length and attenuation. To further
study this dependence, we firstly define the attenuation coefficient ratio $%
\xi =2n\Gamma /c/\alpha _{T}$, who is independent of loop length $L$. And
then the influences of the loop length and attenuation are investigated in
the following.

When the input particle number is $N_{in}=10^{8}$ in a single shot\ and the
gain is $G_{1}=6$ ($G_{2}<G_{1}$ due to the photon loss of the Raman write
beam), the relation between $(\Delta \Omega )_{\min }$ and $L$ with
different $\xi $\ is shown in Fig.~\ref{fig:Fig3}. As we can see, $(\Delta
\Omega )_{\min }$ has two optimal points. This is the result of the
competition between the enhancement from the loop length and the reduction
from the particle number attenuation. When the loop length is small, the
effect of attenuation is small. $(\Delta \Omega )_{\min }$\ is close to that
in the ideal condition shown in the dark-yellow dash line in Fig.~\ref%
{fig:Fig3}.\ With the increase of the loop length, the attenuation in light
and atom exponentially increases and leads to the first optimal point. And
then, $(\Delta \Omega )_{\min }$\ decreases with $L$\ and reaches the second
optimal point. Furthermore, we also give the sensitivity of FOG (see
Appendix) in olive dash line. It is calculated with the same
rotation-sensitive particle number $N_{Tot}$ and the same optical loss as
ALHQG. It can be seen that even with the attenuation, $(\Delta \Omega
)_{\min }$ of the ALHQG at the first optimal point can still beat the SQL ($%
1/(\beta \sqrt{N_{Tot}})$) and is better than the FOG. So we focus on the
parameters of the first optimal point.

\begin{figure}[tbp]
\centering
\includegraphics[scale=0.32]{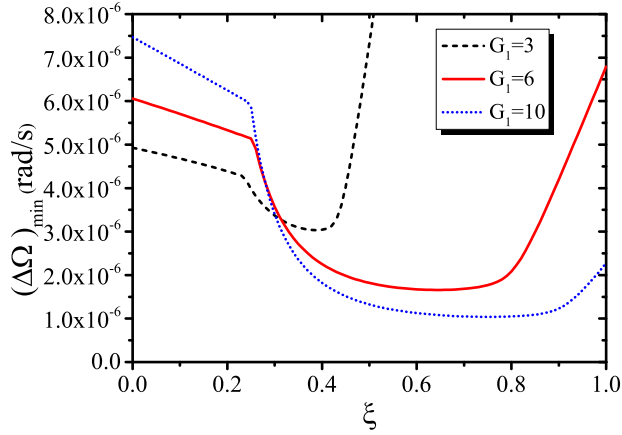}
\caption{The minimum $\Delta \Omega $ versus the attenuation coefficient
rate $\protect\xi$\ with different gains $G_{1}$ when $N_{in}=10^{8}$ in a
single shot\ and $\protect\alpha _{T}=3$ dB/km.}
\label{fig:Fig4}
\end{figure}

Moreover, the minimum rotation sensitivity $(\Delta \Omega )_{\min }$ is
also related with $\xi $. As shown in Fig. \ref{fig:Fig3}, $(\Delta \Omega
)_{\min }$ at $\xi =0.5$\ is worse than that at $\xi =0.7$. When $\xi $ is
given, the minimum rotation sensitivity $(\Delta \Omega )_{\min }$ at the
optimal $L$ can be obtained. The different $\xi $ leads to different $%
(\Delta \Omega )_{\min }$ at the different optimal $L$\ as shown in Fig. \ref%
{fig:Fig4}. Obviously, when $N_{in}=10^{8}$ in a single shot, $G_{1}=6$, $%
\lambda =795$ nm, $D=0.2$ m, $\Lambda (G,N_{in})=0.02286$ rad and $\alpha
_{T}=3$ dB/km, the attenuation coefficient rate should be optimized to $\xi
=0.7$ to get the minimum $\Delta \Omega =2.905\times 10^{-6}$ $rad/s$. In
addition, an important finding is that the optimal $\xi $ increases with
gain $G_{1}$, shown in Fig. \ref{fig:Fig4}. The reason is that there is only
optical input field but no initial atomic spin wave at the input ports of
the ALHQG. The best sensitivity should be achieved at the best interference
visibility. Thus, the intensity balance of two interference arms is
important. That is why the best visibility is always gotten when the
decoherence of atomic beams is smaller than the loss of the optical field,
and the attenuation coefficient rate $\xi $ increases with gain $G_{1}$.
Therefore, a tunable attenuation coefficient rate $\xi $ is really essential
in practice, but this is indicated by few works.

Finally, the ALHQG is built based on the optimal parameters obtained in
above, which are listed as follow: the wavelength $\lambda =795$ nm, the
input particle number in a singel shot $N_{in}=10^{8}$, the gain $G_{1}=6$,
the diameter of the Sagnac loop $D=0.2$ m, the attenuation coefficient in
optical fiber $\alpha _{T}=3$ dB/km, the attenuation coefficient ratio $\xi
=0.7$ and the loop length $L=520$ m. Based on this setup, the optimal
rotation rate $\Omega _{opt}$ $[=\lambda c\Lambda (G_{1},N_{in})/2\pi DL]$
is determined. Using the above set of parameters, Fig. \ref{fig:Fig5} shows
a period of the dynamic range of ALHQG and FOG (see Appendix). The ALHQG can
beat SQL nearby the optimal rotation rate, while FOG cannot. Furthermore,
compared with the earth rotation rate $\Omega _{e}=7.29\times 10^{-5}$
rad/s,\ both ALHQG and FOG cannot work well in all rotation rate. The reason
is that when the rotation rate deviates from the optimal rotation rate, the
increasing of intensity noise leads to the degradation of the rotational
sensitivity for both ALHQG and FOG. Our proposed scheme can further increase
the measured particle number to improve the rotation sensitivity, so that it
can be better than the earth rotation rate in most dynamic ranges except for
the divergence points and its vicinity.
\begin{figure}[tbp]
\centering
\includegraphics[scale=0.32]{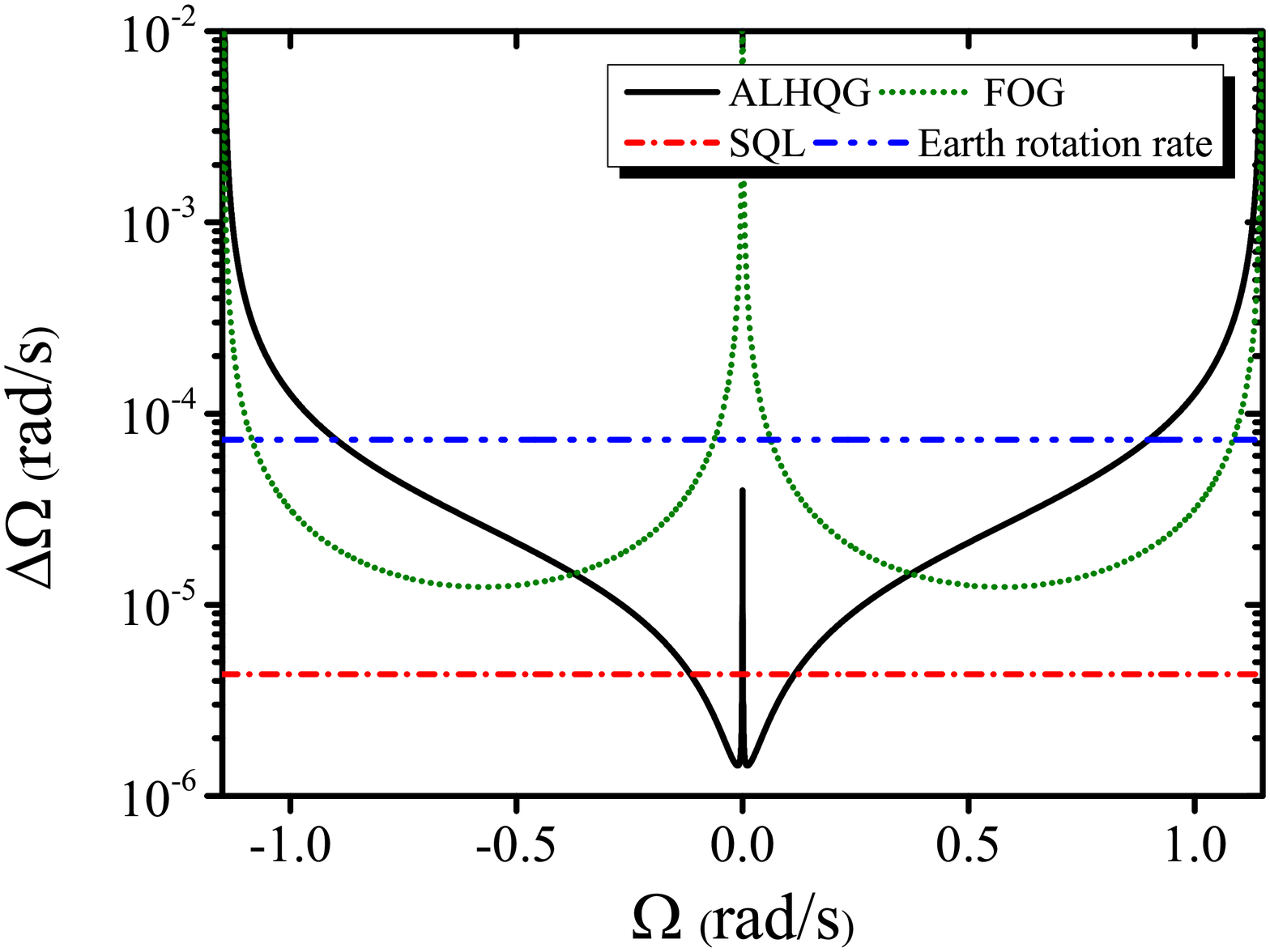}
\caption{Dynamic range. Parameters: $N_{in}=10^{8}$ in a single shot, $%
G_{1}=6$, $\protect\lambda =795$ nm, $D=0.2$ m, $\protect\alpha _{T}=3$
dB/km, $\protect\xi =0.7$ and $L=520$ m. FOG has the same loop length and
the same rotation-sensitive particle number.}
\label{fig:Fig5}
\end{figure}

Therefore, we analyze the dependence of the rotation sensitivity $\Delta
\Omega $ on six parameters, $N_{in}$, $G_{1}$, $L$, $\Omega $, $\alpha _{T}$
and $\Gamma $. In general, $\Delta \Omega $\ decreases with $N_{in}$ and $%
G_{1}$ and increases with $\alpha _{T}$ and $\Gamma $. When $N_{in}$, $G_{1}$%
, $\alpha _{T}$ and $\Gamma $\ are given, the minimum $\Delta \Omega $ can
be obtained\ with the optimal fiber loop length $L$.

\section{Discussion \label{sec4}}

The theoretical analysis based on real experimental parameters is important
to guide the future experimental realization. Now we present the
experimental parameters to obtain the sensitivity of ALHQG \cite{18}. The
Raman amplification process based on $^{87}Rb$ ensemble is employed to
realize the QBS/C. The wavelength of the lights is $\lambda =795$ nm and the
attenuation coefficient in optical fiber is $\alpha _{T}=3$ dB/km. The input
Stokes field is a $1$ $\mu s$ -long pulse with the repetition rate of $10$
kHz and the power of $30$ $\mu W$. Hence, there are $10^{8}$ photons in a
single shot. When the gain is $G_{1}=6$, the diameter of the Sagnac loop is $%
D=0.2$ m, the attenuation coefficient ratio is $\xi =0.7$ and the length of
the Sagnac loop is $L=520$ m, the minimum sensitivity is $2.905\times
10^{-8} $ rad/s/$\sqrt{\text{Hz}}$.

In practical measurement, the sensitivity of the ALHQG may suffer from the
instability of the laser, which mainly has influence on $N_{in}$ and $%
G=\cosh (\left\vert \eta A_{p}\right\vert t)$ where $\eta \varpropto \Delta
^{-1}$. The gain $G$ depends on the amplitude $A_{p}$ and the frequency
detuning $\Delta $ of the strong Raman write beam. Based on Eq. (\ref{RS}),
the fluctuations of $N_{in}$ and $G$ affect the rotation sensitivity.
Normally, the intensity fluctuation of the laser beam can be easily
stabilised within $\pm 0.1\%$. And thus, the fluctuation of $N_{in}$ causes
the fluctuation of the rotation sensitivity between $2.904\sim 2.907\times
10^{-8}$ rad/s/$\sqrt{\text{Hz}}$. Furthermore, the frequency detuning $%
\Delta $\ is about $1$ GHz and its fluctuation is about $1$ MHz. Thus, the
intensity and the frequency fluctuation of the Raman write beam $A_{p}$ are
both within $\pm 0.1\%$, the corresponding gain $G$ fluctuates between $%
5.9978$ and $6.0272$. The rotation sensitivity fluctuates between $%
2.903\times 10^{-8}$ rad/s/$\sqrt{\text{Hz}}$ and $2.907\times 10^{-8}$
rad/s/$\sqrt{\text{Hz}}$. It can be seen that the impact of the fluctuation
of laser on the rotation sensitivity of ALHQG is smaller than $10^{-10}$
rad/s/$\sqrt{\text{Hz}}$.

\section{Conclusion \label{sec5}}

We have proposed an ALHQG, where an atomic ensemble as QBS/C, and an optical
Sagnac loop to couple the rotation rate. Under ideal condition, the value
of the rotation sensitivity, decreasing monotonically with the Raman gain
and Sagnac loop length, can beat the SQL because of the enhancement by Raman
amplification. In the presence of the attenuation of the optical and atomic
interference arms, the sensitivity has two optimal points as the Sagnac loop
length increases. This is the result of the competition between the
enhancement from the loop length and the reduction from the particle number
attenuation. At the first optimal point, the sensitivity can still beat the
SQL. Furthermore, the minimum sensitivity of the ALHQG always surpasses that
of the FOG with the same loop length and the same rotation-sensitive
particle number.

The ALHQG can be operated without complicated phase-locking. Compared with
other kinds of gyroscopes, our gyroscope has advantages of simple structure,
easy operation, and good sensitivity below the SQL. The theoretical analysis
with practically experimental parameters shows that the sensitivity can
reach $10^{-8}$ rad/s/$\sqrt{\text{Hz}}$, which is one order of magnitude
better than that of the FOG. In future, the sensitivity can be improved
further with larger input particle number, larger gain by increasing the
optical depth of the atomic vapor cell. This ALHQG could find practical
application in modern inertial navigation systems.

\section*{Acknowledgments}

We acknowledge financial support from the Natural Science Foundation of
China (Grant No. 11874152, NO. 11904227, No. 11974111, No. 11654005), the
Sailing Program of Shanghai Science and Technology Committee under Grant
19YF1414300, 19YF1421800, Shanghai Municipal Science and Technology Major
Project (Grant No. 2019SHZDZX01), the Fundamental Research Funds for the
Central Universities of China, and the Startup Fund for Youngman Research at
SJTU.

\section*{APPENDIX : THE\ SENSITIVITY ANANLYSIS OF THE FIBER OPTIC GYROSCOPE}

In FOG \cite{10}, a coherent state enters into one port of the gyroscope,
while a vacuum state enters into the other port. After passing through a 3dB
coupler, they propagate in the opposite directions inside a fiber Sagnac
loop. As a result, the lights in CW and CCW experience a Sagnac phase
induced by the rotation rate $\Omega $. Hence, the final photon number is:
\begin{equation*}
\left\langle n\right\rangle _{FOG}=\frac{1}{2}T\left\vert \rho \right\vert
^{2}[1+\cos (\beta \Omega )].
\end{equation*}%
Here $\left\vert \rho \right\vert ^{2}$ is the particle number of the input
field in a single shot. $T=\exp (-\alpha _{T}L)$ is the transmission
coefficient of the photons, where $\alpha _{T}$ is the attenuation
coefficient. $\beta \Omega $ is the Sagnac phase, where $\beta =2\pi
DL/(\lambda c)$, $D$\ is the diameter of the Sagnac loop, $L$ is the length
of the Sagnac loop and $c$\ is the speed of light in vacuum. Therefore,
based on the error-propagation analysis \cite{20}, the rotation sensitivity
of FOG in a single shot is:
\begin{equation*}
(\Delta \Omega )_{FOG}=\frac{1}{\frac{1}{2}T\left\vert \sin (\beta \Omega
)\right\vert \beta \sqrt{\left\vert \rho \right\vert ^{2}}}\propto \frac{1}{%
\beta \sqrt{\left\vert \rho \right\vert ^{2}}}
\end{equation*}%
With the same rotation-sensitive photon number $N_{Tot}=\left\vert \rho
\right\vert ^{2}$ and the same optical loss $T$ as ALHQG, the rotation
sensitivity can be calculated to compare with ALHQG and SQL.


\begin{thebibliography}{99}
\bibitem{1} A. Lawrence, Modern Inertial Technology, Springer, New York
(1998).

\bibitem{2} G. Sagnac, L'ether lumineux demontre par l'effect du vent
relatif d'ether dans un interferometre en rotaion uniforme, C. R. Acad. Sci.
\textbf{157}, 708 (1913).

\bibitem{3} R. Anderson, H. R. Bilger, G. E. Stedman, "Sagnac effect": A
century of Earth-rotated interferometers, Am. J. Phys. \textbf{62}, 975
(1994).

\bibitem{4} A. Kolkiran and G. S. Agarwal, Heisenberg limited Sagnac
interferometry, Opt. Express, \textbf{15} 6798 (2007).

\bibitem{5} B. Barrett, R. Geiger, I. Dutta, M. Meunier, B. Canuel, A.
Gauguet, P. Bouyer, A. Landragin, The Sagnac effects:20 years of development
in matter-wave interferometry, C. R. Physique 16, 343 (2015).

\bibitem{6} Jonathan P. Dowling, Correlated input-port, matter-wave
interferometer: Quantum-noise limits to the atom-laser gyroscope, Phys. Rev.
A \textbf{57}, 4736 (1998).

\bibitem{7} T. L. Gustavson, P. Bouyer, and M. A. Kasevich, Precision
Rotation Measurement with an Atom Interferometer Gyroscope, Phys. Rev. Lett.
\textbf{78}, 2046 (1997).

\bibitem{8} D. Savoie, M. Altorio, B. Fang, L.A. Sidorenkov, R. Geiger, A.
Landragin, Interleaved atom interferometry for high-sensitivity inertial
measurements, Sci. Adv. \textbf{4}, 7948 (2018).

\bibitem{9} I. Dutta, D. Savoie, B. Fang, B. Venon, C. L. Garrido Alzar, R.
Geiger, and A. Landragin, Continuous Cold-atom Inertial Sensor with 1
nrad/sec Rotation Stability, Phys. Rev. Lett. \textbf{116}, 183003 (2016).

\bibitem{10} J. Nayak, Fiber-optic gyroscope: from design to production
[Invited], Appl. Opt. \textbf{50}, E152 (2011).

\bibitem{11} H. C. Lefevre, The Fiber-Optic Gyroscope, Artech House (2014).

\bibitem{12} J. Lautier, L. Volodimer, T. Hardin, S. Merlet, M. Lours, F.
Pereira Dos Santos, and A. Landragin, Hybridizing matter-wave and classical
accelerometers, Appl. Phys. Lett. 105, 144102 (2014).

\bibitem{13} Pierrick Cheiney, Lauriane Fouche, Simon Templier, Fabien
Napolitano, Baptiste Battelier, Philippe Bouyer, and Brynle Barrett,
Navigation-Compatible Hybrid Quantum Accelerometer Using a Kalman Filter,
Phys. Rev. Appl. 10, 034030 (2018).

\bibitem{14} F. Zimmer and M. Fleischhauer, Sagnac interferometry based on
ultraslow polaritons in cold atomic vapors, Phys. Rev. Lett. \textbf{92},
253201 (2004).

\bibitem{15} F. E. Zimmer and M. Fleischhauer, Quantum sensitivity limit of
a Sagnac hybrid interferometer based on slow-light propagation in ultracold
gases, Phys. Rev. A \textbf{74}, 063609 (2006).

\bibitem{16} J. Xin, J. Liu, and J. Jing, Nonlinear Sagnac interferometer
based on the four-wave mixing process, Opt. Express \textbf{25}, 1350 (2017).

\bibitem{17} A. M. Marino, N. V. Corzo Trejo, and P. D. Lett, Effect of
losses on the performance of an SU(1,1) interferometer, Phy. Rev. A \textbf{%
86}, 023844 (2012).

\bibitem{18} B. Chen, C. Qiu, S. Chen, J. Guo, L. Q. Chen, Z. Y. Ou, and W.
Zhang, Atom-light hybrid interferometer, Phys. Rev. Lett. \textbf{115},
043602 (2015).

\bibitem{19} Z. D. Chen, C. H. Yuan, H. Ma, D. Li, L. Q. Chen, Z. Y. Ou, and
W. Zhang, Effects of losses in the atom-light hybrid SU(1,1) interferometer,
Opt. Express \textbf{24}, 17766 (2016).

\bibitem{20} Vittorio Giovannetti, Seth Lloyd, Lorenzo Maccone,
Quantum-Enhanced Measurements: Beating the Standard Quantum Limit, Science
306, 1330 (2004).
\end{thebibliography}
\end{document}